\documentclass[aps,pra,twocolumn,superscriptaddress]{revtex4-2}
\usepackage{amssymb}
\usepackage{amsmath}
\usepackage{amsthm}
\usepackage{xfrac}
\usepackage{bm}
\usepackage{graphicx}
\usepackage{subcaption}
\usepackage{ragged2e}
\usepackage{float}
\usepackage[hyperfootnotes=false]{hyperref}
\usepackage{tabularx}
\usepackage{dcolumn}
\usepackage{isotope}
\usepackage[dvipsnames,svgnames]{xcolor}
\usepackage{siunitx}
\usepackage{bbold}

\DeclareMathOperator{\sinc}{sinc}
\DeclareMathOperator{\rect}{rect}

\colorlet{MAGENTA}{magenta}

\theoremstyle{definition}

\newtheorem{benchmark}{Benchmark}

\usepackage{subcaption}
\usepackage[export]{adjustbox}
\begin{document}

\title{Optimizing two-qubit gates for ultracold fermions in optical lattices}

\author{Jan~A.~P.~Reuter\,\href{https://orcid.org/0009-0008-6786-4053}{\includegraphics[height=6pt]{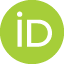}}}
\email[]{ja.reuter@fz-juelich.de}
\affiliation{Forschungszentrum Jülich GmbH, Peter Grünberg Institute, Quantum Control (PGI-8), 52425 Jülich, Germany}
\affiliation{Institute for Theoretical Physics, University of Cologne, Zülpicher Straße 77, 50937 Cologne, Germany}

\author{Juhi~Singh\,\href{https://orcid.org/0000-0001-9807-9551}{\includegraphics[height=6pt]{ORCID-iD_icon-64x64.png}}}
\affiliation{Forschungszentrum Jülich GmbH, Peter Grünberg Institute, Quantum Control (PGI-8), 52425 Jülich, Germany}
\affiliation{Institute for Theoretical Physics, University of Cologne, Zülpicher Straße 77, 50937 Cologne, Germany}

\author{Tommaso~Calarco\,\href{https://orcid.org/0000-0001-5364-7316}{\includegraphics[height=6pt]{ORCID-iD_icon-64x64.png}}}
\affiliation{Forschungszentrum Jülich GmbH, Peter Grünberg Institute, Quantum Control (PGI-8), 52425 Jülich, Germany}
\affiliation{Institute for Theoretical Physics, University of Cologne, Zülpicher Straße 77, 50937 Cologne, Germany}
\affiliation{Dipartimento di Fisica e Astronomia, Università di Bologna, 40127 Bologna, Italy}

\author{Felix~Motzoi\,\href{https://orcid.org/0000-0003-4756-5976}{\includegraphics[height=6pt]{ORCID-iD_icon-64x64.png}}}
\affiliation{Forschungszentrum Jülich GmbH, Peter Grünberg Institute, Quantum Control (PGI-8), 52425 Jülich, Germany}
\affiliation{Institute for Theoretical Physics, University of Cologne, Zülpicher Straße 77, 50937 Cologne, Germany}

\author{Robert~Zeier\,\href{https://orcid.org/0000-0002-2929-612X}{\includegraphics[height=6pt]{ORCID-iD_icon-64x64.png}}}
\email[]{r.zeier@fz-juelich.de}
\affiliation{Forschungszentrum Jülich GmbH, Peter Grünberg Institute, Quantum Control (PGI-8), 52425 Jülich, Germany}

\date{March 27, 2026}

\begin{abstract}
Ultracold neutral atoms in optical lattices are a promising platform for simulating the behavior of complex materials and implementing quantum gates. We optimize collision gates for fermionic \isotope[6]{Li} atoms confined in a double-well potential, controlling
the laser amplitude and keeping its relative phase constant.
We obtain high-fidelity gates based on a one-dimensional confinement simulation.
Our approach extends beyond earlier Fermi-Hubbard simulations
by capturing a momentum dependence in the interaction energy.
This leads to
a higher interaction strength
when atoms begin in separate subwells compared to the same subwell.
This momentum dependence might limit the gate fidelity under realistic experimental conditions, 
but also enables tailored applications in quantum chemistry and quantum simulation 
by optimizing gates for each of these cases separately.
\end{abstract}

\maketitle

\section{Introduction\label{Introduction}}
Quantum computation with neutral atoms in optical lattices has seen rapid progress due to the ability to scale to large atom numbers and to perform high-fidelity local operations \cite{Bluvstein2024, Evered2023, Gyger2024, Shaw2024}. A key challenge, however, remains the realization of fast and high-fidelity two-qubit gates that are compatible with scalable architectures. One promising approach is the use of collisional gates, where entanglement is generated through interactions when atoms are brought into contact \cite{Busch1998,Calarco2000,Idziaszek2005,Liang2008}. This method leverages the s-wave scattering \cite{Bloch2008,Sakurai2020} interaction between ultracold atoms and is inherently robust to certain types of noise.
The concept of using controlled collisions in optical lattices for quantum gates was first proposed by \cite{Jaksch1975}, where the interaction-induced phase shift from s-wave scattering was used to implement entangling gates between atoms in adjacent lattice sites. Subsequent experiments with bosonic atoms, such as 
\isotope[87]{Rb} in double-well potentials \cite{Mandel2003,Calarco2004,Anderlini2007,Trotzky2008,Yang2020,Impertro2024}, have demonstrated coherent two-body dynamics and entanglement generation via controlled tunneling and collisions.
Recent attention has turned to fermionic systems \cite{Zöller2011,Murman2015,Hartke2022,Zhu2025,Xu2025,Tabares2025,Mark2025,Kiefer2025}, especially using atoms as
\isotope[6]{Li}, due to their favorable collisional properties, large Feshbach resonance tunability, and small mass which enables faster tunneling. Fermions offer additional advantages for quantum information processing, such as spin-encoded qubits \cite{Gonzalez2023} and Pauli-blockade induced noise protection \cite{Sanner2021}. Theoretical works as well as experiments have demonstrated local control over spin and position states of fermionic atoms in optical tweezers and lattices \cite{Hauck2022,Cicali2024,Chalopin2025,Bojovic2025}.

With precise control of the system, efficient single- and two-qubit operations are achieved. Recent works primarily employ 
different types of controls to steer the dynamics of the system and perform gates. Using a relative phase between the lasers, single-qubit operations, specifically $\pi$-pulses, have been performed with an approximate fidelity of 0.993 and a pulse duration of about $\SI{500}{\micro\second}$ \cite{Chalopin2025}. The longer pulse duration and its shape have been chosen to suppress excitations of the atom. The residual error arises from various noise sources, such as phase and intensity noise. 
Adjusting the amplitude of the lasers to change the barrier height between the wells, a recent experiment \cite{Bojovic2025} performed
a two-qubit entangling operation with a fidelity of 0.9975 and a gate duration of $\SI{1.2}{\milli\second}$. The gate sequence is designed to balance gate speed while limiting doublon creation.

These quantum operations have been performed with great precision, the gate durations and fidelities can be further improved using optimal control techniques. 
In this work, we build on earlier work and tools as in \cite{Nemirovsky2021,Singh2025} to optimize 
fermionic two-qubit entangling gates in a double well based on a one-dimensional (1D) confinement simulation, while controlling
the laser amplitude and keeping its relative phase constant.
To this end, we develop an alternative numerical simulation technique (see Sec.~\ref{leapfrog}) 
which goes beyond the Fermi-Hubbard simulation in \cite{Singh2025}.
This simulation method is reasonably fast and allows for adequate precision.
We benchmark it against the methods in \cite{Nemirovsky2021,Singh2025}
(see Sec.~\ref{Bench marking the simulation method})
using realistic parameters from the recent experimental work \cite{Chalopin2025}.
Our simulation method is then combined with an
innovative gradient-based optimization (see Sec.~\ref{Gradient based optimization}).

Based on the developed simulation and optimization methods,
we observe for a unitary gate optimization in Sec.~\ref{Unitary gate optimization} characteristic differences compared to earlier work.
Exploring this difference in Sec.~\ref{Case optimization}, the 1D confinement simulation highlights a momentum dependence of the interaction energy, which results in a higher interaction strength for atoms starting in different subwells compared to atoms starting in the same subwell.
This is not directly observable in the Fermi-Hubbard model
as it does not consider the momentum information.
The 1D confinement simulation enables us to optimize the two cases separately (see Sec.~\ref{Case optimization}) which is relevant
for important applications in quantum chemistry \cite{Gkritsis2025} (starting from the same subwell)
and in the context of quantum simulation and quantum computing \cite{Chalopin2025,Singh2025,Bojovic2025}
(starting in separate subwells).
Finally, our optimized gates are robust against an asymmetric lattice due to
a relative phase in the laser control and 
an error in the interaction energy which occurs due to uncertainties in the effective interaction
(as shown in Sec.~\ref{Robustness against system impurities}).
Moreover, we analyze how the gates are affected
by state-preparation errors
resulting in three particles in a single double well.

The manuscript introduces its physical system description and its target gate in Sec.~\ref{The system}.
It continues in Sec.~\ref{Simulation and optimization methods} with the developed simulation and optimization
methods and the corresponding results are detailed in Sec.~\ref{Optimization of Vs, Vl and a in the 1D confinement}.
The robustness of our gates is discusses in  Sec.~\ref{Robustness against system impurities} before we conclude.

\section{System description\label{The system}}
\begin{figure}[b]
\begin{subfigure}{0.49\linewidth}
	\includegraphics[width=\linewidth]{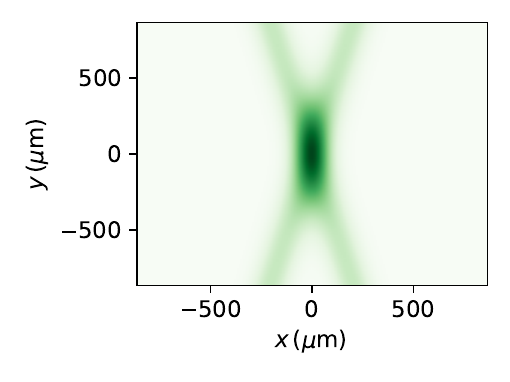}
\end{subfigure}
\begin{subfigure}{0.49\linewidth}
	\includegraphics[width=\linewidth]{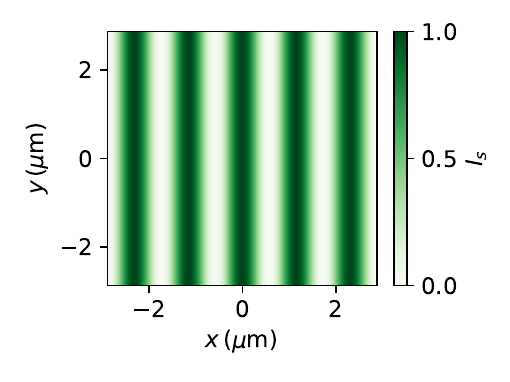}
\end{subfigure}
\caption{\justifying \textbf{Optical lattice from a tilted laser field.} On the left,  intensity of the laser field 
acting on the short lattice
in the $xy${-}plane 
with a characteristic angle of $\beta_{x}=26.7^\circ$
between the two laser beams. On the right, a magnified version details
 the $\cos^2(k_x x)$ lattice.
\label{Lattice}}
\end{figure}

We consider localized ultra{-}cold fermionic \isotope[6]{Li} atoms in a three-dimensional optical lattice with coordinates $\mathbf{r}=(x,y,z)$.
The corresponding optical potential $V(\mathbf{r},t)=V(x,t){+}V(y){+}V(z)$ can be split into its three spatial components.
The y{-} and z{-}components $V(y)= {-}V_y \cos^2 (k_y y)$ and $V(z)= {-}V_z \cos^2 (k_z z)$
are created by single standing waves  and
are kept constant with time.
The $x${-}direction is time dependent and contains two different standing waves
\begin{equation}
V(x,t)= V_s(t) \cos^2(k_x x {+} \phi) - V_l(t) \cos^2 \left(\frac{k_x}{2} x \right).
\label{eq potential}
\end{equation}
Here, $V_s$ and $V_l$ are the short and long lattice potential energies and they represent two of our controls of the system.
The wave numbers of the laser fields are given by
$k_d=\sin({\beta_{d}}/{2})\, {2\pi}/(\SI{532}{\nano\meter})$ for $d \in \{x,y,z\}$,
where the $x$- and $y$-angles are given by $\beta_{x}=\beta_{y}=26.7^\circ$. We discuss in Sec.~\ref{Unitary gate optimization}
both the cases of $\beta_{z}= 26.7^\circ$ and $\beta_{z}=10.16^\circ$, where
the latter case correspond to the angles in \cite{Chalopin2025,Bojovic2025,exp_data}.
The relative phase $\phi$ between the long and short lattice will in this work be set to zero, except for Sec.~\ref{Bench marking the simulation method}.
In Fig.~\ref{Lattice} one can see the resulting lattice in $x$-direction produced by the overlapping laser fields.

\begin{figure}[b]
\includegraphics[width=0.75\linewidth]{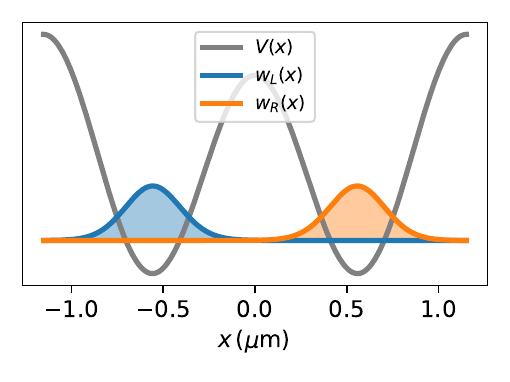}
\caption{\justifying \textbf{The maximally localized Wannier states.} The atoms are prepared in either the left or the right Wannier state. Both states are a superposition of the two lowest Bloch bands of the corresponding optical potential in the $x${-}direction.}
\label{Wannier states}
\end{figure}

Equation~\eqref{eq potential} has been simplified by
neglecting the (approximate)
Gaussian decay $\exp\{{-}2[x\cos({\beta_{x}}/{2})/{\mathrm{w}_s}]^2\}$
of the laser intensity in the $x${-}direction. We assume a
laser with a Gaussian beam with beam waists of 
$\mathrm{w}_s \approx \SI{120}{\micro \meter}$ and $\mathrm{w}_l \approx \SI{220}{\micro \meter}$.
Localized atoms in an optical lattice are considered
 to be in a maximally localized Wannier state $w(x)$. This state is an equal superposition
of all Bloch states $e^{i qx}u_q(x)$ of that lattice $w(x)=\sum_q e^{i (qx{+}\chi_q)}u_q(x)/\sqrt{N}$ with a momentum depending phase $\chi_q$ and 
an arbitrary number $N$ of sites in the lattice.
We obtain the value of $\chi_q$ by 
solving the equation $\mathrm{x} w(x{-}x_0) =x_0 w(x{-}x_0)$, where
$\mathrm{x}$ is the position operator in the Bloch basis and $x_0$ is the expectation value of the position.
The double-well structure in the $x${-}direction yields
a symmetric (even) and an antisymmetric (odd) solution with a small band gap.
The left ($w_L$) or right ($w_R$) localized Wannier states are then a
superposition between the even $2n$ band and its subsequent odd $2n{+}1$ one
\cite{Wannier1937,Kohn1959,Kohn1973,Singh2025}.
 An example for the two lowest bands ($n=0$) are shown in Fig.~\ref{Wannier states}.

We now move from one to two atoms and aim at entangling two atoms with each other by applying a collision gate in the $x$-direction. The atoms sit both in one of the double wells created by $V(x,t)$.
For the two{-}body Hamiltonian with coordinates $\mathbf{r}_j=(x_j,y_j,z_j)$, a pseudo potential is given by the $\delta$-function
which represents the $s$-wave scattering and results in
\begin{align}
H(\mathbf{r}_1,\mathbf{r}_2,s_1,s_2,t) = & \sum_{j=1}^2 {-}\frac{\hbar^2}{2m} \nabla_j^2{+}V(\mathbf{r}_j,t) \notag \\
& + U_{3\mathrm{D}} \delta_{s_1,\bar{s}_2}\delta_{\mathrm{reg}}(\mathbf{r}_1{-}\mathbf{r}_2)
\label{eq Hamiltonian}
\end{align}
where $s_j \in \{ {\uparrow,\downarrow} \}$ is the spin of the atom $j$ with mass $m$ (and $\bar{s}_j=\,\,\downarrow$ for $s_j=\,\,\uparrow$ and vice versa),  $U_{3\mathrm{D}}={4 \pi \hbar^2 a}/m$ \cite{Bloch2008}, $a$ denotes the scattering length, and $\delta_{\mathrm{reg}}$ is the regularized delta function \cite{Busch1998,Idziaszek2005,Liang2008}.

\subsection{The \texorpdfstring{$s$}{s}-wave scattering in the 1D confinement\label{sec:s-wave}}
As introduced above, the wave function of one atom has three geometric dimensions $x$, $y$, and $z$. Consequently, 
the wave function of two atoms is defined by the six-dimensional geometric space given by the coordinates $\mathbf{r}_j=(x_j,y_j,z_j)$.
Or in general, the number of dimensions $\mathrm{D}$ of a combined wave function is the sum 
$\mathrm{D}=\sum_j \mathrm{D}_j$ of the dimensions $\mathrm{D}_j$ of the separated wave functions.
Since the simulation time for such a problem grows exponential with the number of dimensions (see Sec.~\ref{leapfrog}), we truncate
the wave function. If one deals with a nearly harmonic potential, one can approximately neglect the ``center of mass" part, 
e.g., $y_1{+}y_2$, of the wave function based on its rotation symmetry as it is mostly decoupled from the distance part,
e.g., $y_1{-}y_2$.
Alternatively, one can effectively describe each atom in lower confinements to truncate the wave function.

We consider the wave functions to be strongly confined in the $y${-} and $z${-}directions due to 
our assumption of a strong and narrow potential (similarly as in \cite{Nemirovsky2021}).
This allows us to neglect the dynamics in these directions
during the collision, which is denoted as the 1D confinement since only one spatial degree of freedom remains.
Thus confinement-induced resonances can occur with an effective 1D scattering potential
$U_{1\mathrm{D}}\delta (x_1{-}x_2)$ \cite{Bloch2008, Hadzibabic2011, Haller2010, Lorenzi2023},
\begin{equation}
U_{1\mathrm{D}}={-}\frac{2 \hbar^2}{m a_{1\mathrm{D}}} \;\text{and}\;
a_{1\mathrm{D}}=l_\perp \left(A{-}\frac{l_\perp}{a}\right) , 
\label{U1D}
\end{equation}
where $l_\perp$ is the oscillation length in the $yz${-}plane and $A=1.036$ has been numerically calculated \cite{Bloch2008,Bergman2003,Petrov2001}. 

\subsection{Two{-}atom Wannier states in the 1D confinement\label{Two{-}atom Wannier states in the 1D confinement}}
We have assumed until now that our atoms will be prepared in an initial state
that is approximately a Wannier state of the optical lattice in Eq.~\eqref{eq potential}
while neglecting any interaction between them. If, however, both atoms are localized on the same side of the double well, 
they will strongly interact with each other and the optical lattice alone can no longer define our initial state.
Consequently, we need to define new two{-}atom Wannier states, which take also the interaction energy into account.
The interaction potential is however not periodic and thus can not lead to Wannier states (as they need to be periodic, see the beginning of Sec.~\ref{The system}).
We can resolve this by using a Dirac comb $\sum_{n={-}\infty}^{\infty} \delta[x_1{-}x_2{-}n{2\pi}/(\sqrt{2}k_x)]$
in addition to the optical lattice, instead of the single term corresponding to $n=0$ in the sum.
This results in a periodicity of $\sqrt{2}k_x$, which is big enough so that its effect on the Wannier state is approximately only given for $n=0$.

\begin{figure}[t]
\includegraphics[]{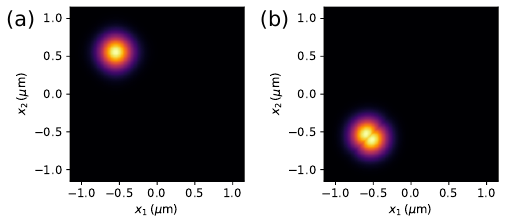}
\caption{\justifying \textbf{Wannier states of interacting atoms.} (a)~shows the decoupled two{-}atom Wannier state with the first atom in the left 
state $w_L(x_1)$ and the second one in the right  state $w_R(x_2)$. (b)~shows the coupled two{-}atom Wannier state $w_{LL}(x_1,x_2)$, where the interaction
energy at $x_1=x_2$ divides the wave packet in two symmetric parts. We have recoil energies $V_s=40 E_{r,s}$, $V_l=30 E_{r,l}$,
and an effective scattering length of $a_{1\mathrm{D}}=-6675 a_0$ with Bohr radius $a_0 \approx \SI{52.9}{\pico \meter}$.
\label{Basis Wannier states}}
\end{figure}

In general, this holds for all four possible two{-}atom positions $LL$, $LR$, $RL$, $RR$.
If both particles sit in different subwells, e.g., $LR$, their Wannier states can 
nearly perfectly described by the multiplication of the Wannier states of the separated atoms $w_{LR}(x_1,x_2)\approx w_L(x_1)w_R(x_2)$.
In contrast to that, we need to calculate the new combined Wannier states
for the cases where both particles sit in the same subwell [e.g., $LL$] by adapting to the interaction. These new Wannier states 
then represent already entangled particles since the wave packet is divided in two symmetric parts [see Fig.~\ref{Basis Wannier states}].

\subsection{The quantum gate \label{sec:quantum:gate:optimization}}
We discuss the quantum gate that will be applied to the atoms.
The potential energy of the laser field is chosen such that its values at the beginning and the end of the
gate agree. Thus we can use the same Wannier basis to describe the initial and the target state
and this Wannier basis represents our computational basis.

We start with the two associated Wannier states in the $x${-}direction $\vert L \rangle:=w_L (x)$, $\vert R \rangle:=w_R (x)$
which are localized in the left or right subwell.  We then define
the operators $\hat{X}=\vert R \rangle \langle L \vert {+} \vert L \rangle \langle R \vert$ and $\hat{Z} = \vert L \rangle \langle L \vert {-} \vert R \rangle \langle R \vert$ and 
this allows us to describe the single-atom target gate as a partial rotation around $-\hat{X}$ by an angle $\alpha$, i.e.,
\begin{equation}\label{eq:P:T:1}
\hat{P}_{T,1}(\alpha)=\exp \left[-i \frac{\alpha}{2} (-\hat{X}) \right].
\end{equation}
The minus sign in $-\hat{X}$ is arbitrary and originates from the fact that we choose $w_R(x)=w_L(-x)$.
For two atoms, one first defines an operator 
 $\hat{\delta}={(\hat{\mathbb{1}} \otimes \hat{\mathbb{1}} {+} \hat{Z} \otimes \hat{Z})}/{2}$
representing the interaction in our
computational basis, which is possible as we consider only interactions on
a single side of the double well.
Thus we consider
\begin{equation*}
\hat{\mathcal{H}}={-} \frac{\sqrt{3}}{4} (\hat{X} \otimes \hat{\mathbb{1}} {+} \hat{\mathbb{1}} \otimes \hat{X}) {+} \hat{\delta}
\end{equation*}
which resembles the two-band Hubbard-model Hamiltonian.
To maximally entangle two atoms starting with one atom sitting on the left $\vert L \rangle$ and the other one on the right $\vert R \rangle$,
one chooses $U/J={4}/{\sqrt{3}}$ \cite{Yang2020, Trotzky2008} as in the Hubbard model.
We can define a partial transition gate for two atoms as
\begin{equation}\label{eq:P:T:2}
\hat{P}_{T,2}(\alpha)=\exp({-}i \alpha \hat{\mathcal{H}}).
\end{equation}
In each case, our gate takes only the position of the atoms into account and assumes fixed, opposite spins. 
To decompose the total quantum state into a position-spin basis, 
one has to respect the Pauli principle and needs to use the Slater determinant.
Table~\ref{table:A} details a look{-}up table with all 16 possible cases in a double
well for zero to four atoms in the lowest band.

\begin{figure}[b]
\includegraphics[width=0.65\linewidth]{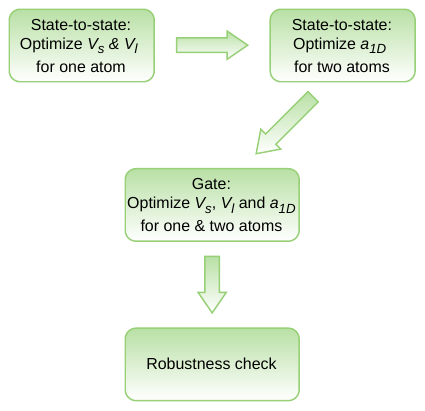}
\caption{\justifying \textbf{The optimization process as a flow chart.} In the first three steps, we optimize the laser intensities $V_s$, $V_l$ and the effective scattering length $a_{1\mathrm{D}}$.
In the last step, we check of the robustness of the gate optimization, including experimental imperfections such as an asymmetric lattice and uncertainties in the interaction energy or the laser intensity (see Sec.~\ref{Robustness against system impurities}).
This optimization procedure is tailored to minimize the total time of the optimization (for details see Sec.~\ref{Optimization of Vs, Vl and a in the 1D confinement}).
\label{Process}}
\end{figure}

We summarize the complete optimization procedure of Sec.~\ref{Optimization of Vs, Vl and a in the 1D confinement} in Fig.~\ref{Process}. 
This optimization procedure is tailored to minimize the total time of the optimization.
We first optimize the time-dependent intensity of the short $V_s(t)$ and long $V_l(t)$ lattice for a single particle
(which has by far the lowest computation time).
Using the optimized $V_s(t)$ and $V_l(t)$ and keeping them unchanged,
the gate is then optimized in the two-particle subspace (which has an intermediate computation time) based on the 1D-confinement approach by varying the time{-}constant effective scattering length $a_{1\mathrm{D}}$.
After these state-to-state optimizations,
we simultaneously optimize $V_s(t)$, $V_l(t)$, and $a_{1\mathrm{D}}$
in the last step 
via a gate optimization (which is computationally the most expensive step)
using the results from the first two preoptimizations as an initial guess.
Finally, we will test the robustness of the gate optimization
in Sec.~\ref{Robustness against system impurities}.

\section{Simulation and optimization methods\label{Simulation and optimization methods}}
\subsection{Numerical time evolution\label{Numerical time evolution}}
A common approach to solve the Schrödinger equation for a continuous wave function is to use a split{-}step method \cite{Sinkin2003,Nemirovsky2021}, where we switch between the real and the momentum space. This method relies on a first order Trotterization of the time evolution operator and uses a fast Fourier transform (FFT) $\mathcal{F}$  to diagonalize the kinetic energy part of the Hamiltonian \cite{Montangero2018, Bao2002, Speth2013}: 
\begin{equation}
e^{{-}i H(x,t) \frac{dt}{\hbar}}\approx e^{{-}i V(x,t) \frac{dt}{2\hbar}} \mathcal{F}^\dagger e^{{-}i\frac{(\hbar q)^2}{2m} \frac{dt}{\hbar}} \mathcal{F} e^{{-}i V(x,t) \frac{dt}{2\hbar}}.
\end{equation}

The FFT acts on the whole wave function to provide global information about the momentum, instead of using a discrete
local approximation of the second derivative. Thus one only needs a few grid points in the real and momentum space for a good approximation of the
time evolution. But the computation time of the parallelized FFT is still the largest
part of the simulation time and is given by $\mathcal{O}(N \log (N))$, where $N$ is the number of grid points.
Since $N$ grows in general exponentially with the number of spatial dimensions of the wave function, the computation time increases
very unfavourably. We aim for a faster approach, as we will simulate many iterations of our two atom-gate, each with a
different set of controls, to find the one which gives us the best result.

\subsubsection{Alternative ``leapfrog" method \label{leapfrog}}
We will now introduce an approach for which the computation time grows approximately linearly with $N$.
We use a Numerov-type ansatz \cite{Numerov1927} which is applied to the time instead of 
a spacial dimension and where we approximate the total differential of the wave function in time with
\begin{equation}
\psi(t{+} \tfrac{dt}{2}){-}\psi(t{-}\tfrac{dt}{2}) \approx {-}\tfrac{idt}{\hbar}H(t)\psi(t)
\label{eq. leapfrog}
\end{equation}
up to first order by inserting the Schrödinger equation for the time derivative. The resulting integration method evokes the leapfrog method \cite{Hockney2021} from classical mechanics.

For the second derivative in real space given by the Hamiltonian in Eq.~\eqref{eq Hamiltonian}, we choose a discrete third-order approximation with
periodic boundary conditions.
As we assume a periodic potential, one could also perform the simulation in momentum space, which would render
the third-order approximation of the second derivative unnecessary. Nevertheless, we test the simulation method in real space and its applicability to general potentials where this speedup would not be applicable.

\subsubsection{Benchmarking the simulation method\label{Bench marking the simulation method}}
We test our simulation method from Sec.~\ref{leapfrog} and compare it with other methods from prior work \cite{Nemirovsky2021,Singh2025}.

\begin{benchmark}\label{bench:one}
In the first test, we will repeat the procedure of \cite{Chalopin2025} and compare our simulation results, obtained with the ``leapfrog" method, to their experimental data in Fig.~\ref{Benchmark 1}.
We consider an ensemble of non{-}interacting atoms in a 1D superlattice, which is initially prepared in the lowest left{-}localized Wannier state of an
asymmetric double well with non{-}zero relative phase $\phi$. Then the relative phase is changed linearly within the first $\SI{0.05}{\milli \second}$ between the long and the short lattice to initiate the Rabi
oscillations between the left and the right well. To include the effect of atoms hopping to neighboring wells, a single atom is simulated
starting in a central double well, which is surrounded by two other double wells. The total probability of finding an atom in the left subwell is obtained by integrating
the squared absolute value of the wave function over the left half of each double well. We refer to \cite{Chalopin2025}
for a detailed description of the experimental procedure including all of the physical parameters. Additionally, we simulate the same setup with the simulation
method of \cite{Singh2025} to compare it to our ``leapfrog" method. In \cite{Singh2025}, the quantum state and the Hamiltonian 
are described in a Wannier basis and one projects both of them into a new Wannier basis every time the potential changes;
in other words, they use a Fermi-Hubbard simulation with higher bands.
We  validate that our method can capture the physics from the Fermi-Hubbard simulation in \cite{Singh2025}.

\begin{figure}[t]
\includegraphics[width=0.75\linewidth]{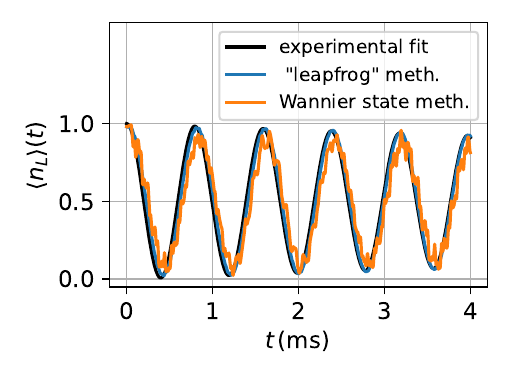}
\caption{\justifying \textbf{Benchmark 1.} Simulated Rabi oscillations of non{-}interacting atoms between the left and the right side of a double-well lattice. The blue line is calculated using Eq.~\eqref{eq. leapfrog} while the orange line uses the method from Ref.~\cite{Singh2025}. The black line corresponds the experimental data \cite{Chalopin2025} which is almost superimposed
with the blue line.
The damping of the oscillation amplitude is caused by atoms hopping out of their initial well.}
\label{Benchmark 1}
\end{figure}

In Fig.~\ref{Benchmark 1}, the simulation result of our ``leapfrog"-method matches almost perfectly with the measured Rabi oscillations of \cite{Chalopin2025} which was given by $\langle n_L \rangle(t)=[1+\cos (\omega t)\exp(-t/\tau_d)]/2$ with $\omega=2\pi {\cdot} \SI{1.261(1)}{\kilo \hertz}$ and $\tau_d=\SI{27(3)}{\milli \second}$.
Also the simulation result using the ``Wannier state"{-}method from \cite{Singh2025} roughly agrees, even though the oscillation shows a jitter which results from the limited amount of basis states.
This is a consequence of how \cite{Chalopin2025} changes the relative phase; our simulation mimics this approach.
A relative phase is applied by changing the short lattice which results
in a shift of the potential minima in the $x$-direction. This shift rapidly increases the number of basis states necessary to represent states
which were originally represented in the unshifted basis. This effect is well known from 
a shifted Gauss packet which is represented as  a coherent state. We discuss in Appendix~\ref{Apprendix A}
the simulation result for the case in which the relative phase is manipulated via the long lattice.

The simulation was performed with potentials energies of $V_s=10.3 E_{r,s}$ and $V_s=31.9 E_{r,l}$, which is in the range of the measured experimental laser intensities \cite{Chalopin2025,exp_data} of $V_s=10(0.5) E_{r,s}$ and $V_s=31(1) E_{r,l}$. 
The computation time for the two simulations have been $\SI{0.19}{\second}{+}\SI{5.87}{\second}$ with our method and $\SI{14.45}{\second}{+}\SI{1.44}{\second}$ with the ``Wannier state"{-}method from \cite{Singh2025}. 
The first number, which is also the more important one, marks the time which was needed to simulate the period where the relative phase has been changed, while the second one marks the time where the potential stays time independent. 
This is not unexpected, because \cite{Singh2025} requires to calculate a new set of Wannier states at each time step in the first part, which is computationally expensive. 
Vice versa, it is cheaper to calculate the time evolution for a small time-independent Hamiltonian in the second part. In comparison, our method does only slightly depend on the change from a time-dependent potential to a stationary one.
In summary, our method is in general more precise and, assuming a lower number of particles or geometric dimensions, it is also faster for time-dependent problems.
\end{benchmark}

\begin{figure}[b]
\includegraphics[]{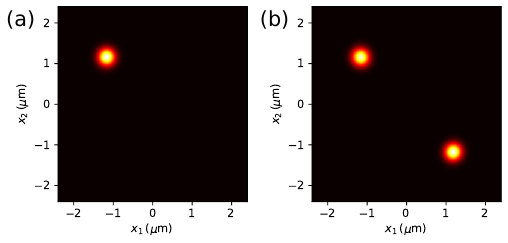}
\caption{\justifying \textbf{Benchmark 2.} An initial Gaussian wave packet $\psi_0$ is placed in a broad centered harmonic trap (left panel). Due to the interaction energy at
$x_1{=}x_2$, the wave packet splits up into two equal Gaussian wave packets with a relative phase of ${-}i$ (right panel).}
\label{Benchmark 2}
\end{figure}

\begin{benchmark}\label{bench:two}
We will now simulate two atoms that have been initially prepared in the ground state of two Gaussian traps. We will consider these traps to be harmonic, which is a good approximation for small deviations from the center of each trap. Consequently, we can write their corresponding ground state as $\psi_0(x)=\exp[{-}{(x\pm {d}/{2})^2}/(2\sigma^2)]/{\sqrt{\sqrt{\pi}\sigma}}$
with a relative distance $d$ between the traps. At time $t{=}0$, we will turn off the initial harmonic traps and turn on a shallower harmonic potential $V(x){=}m \omega^2 x^2/2$
centered between the two atoms with a variance ${\hbar}/(2m \omega) {\gg} \sigma^2/2$, which is much bigger than the one from the initial traps. 
The considered physical parameters are $d{=}\SI{2.329}{\micro \meter}$, $\sigma{=}\SI{0.148}{\micro \meter}$, and $\omega{=}2\pi {\cdot} \SI{43.671}{\mega \hertz}$. The 1D interaction energy is given by the pseudo potential described
in Sec.~\ref{sec:s-wave}. This simulation setup was initially considered in \cite{Nemirovsky2021}, which aims at creating an entangling $\sqrt{\mathrm{SWAP}}$ gate, 
but with fixed harmonic traps instead of a variable double-well potential. We apply the same physical and numerical parameters as in their work. 
This specific problem can now be solved in two ways, either with a 2D simulation or in two 1D simulations, since one can use the rotational symmetry of the Hamiltonian to separate the distance $x_1{-}x_2$ from the center of mass $x_1{+}x_2$. We will simulate both possibilities and compare their fidelity and computation time.
The fidelity of our method is $0.9974$ with a computation time of $\SI{505,64}{\second}$ (2D) and $0.9975$ within $\SI{6,29}{\second}$ ($2\times$1D). The fidelity of the Fourier method is $0.9974$ with a computation time of $\SI{537.73}{\second}$ (2D) and $0.9976$ within $\SI{19,11}{\second}$ ($2\times$1D). As a comparison, \cite{Nemirovsky2021} obtains a fidelity of $0.9979$.
In Figure~\ref{Benchmark 2}, one can see the initial and final state of the simulation for our method in 2D,
which is almost identical to the simulation in \cite{Nemirovsky2021}.
This can be seen as a validation of our simulation method as it yields 
reliable results for time{-}dependent potentials and for more than one simulated dimension.
\end{benchmark}

\subsection{Gradient-based optimization method\label{Gradient based optimization}}
The combined gate infidelity $\epsilon=1{-}(\vert O_1 \vert^2{+}\vert O_2 \vert^2)/{2}$ describes 
how well our optimized gates match with the corresponding target gates [Eqs.~\eqref{eq:P:T:1}-\eqref{eq:P:T:2}], where
\begin{align*}
\vert O_1 \vert(\alpha,\tau) &=\mathrm{Tr}(\vert \hat{P}_{T,1}^\dagger(\alpha) \hat{\Psi}_1(\tau) \vert)/2
\intertext{is used for single atoms and}
\vert O_2 \vert(\alpha,\tau) &=\mathrm{Tr}(\vert \hat{P}_{T,2}^\dagger(\alpha) \hat{\Psi}_2(\tau) \vert)/4
\end{align*} 
for two atoms. Here, $\tau$ is the gate time and the matrices $\hat{\Psi}_i$ are defined below.
In the definition of the unitary gate infidelity $\epsilon$, we chose the equal weight of $1/2$ between the single- and the two-atom gate since we will apply our pulses globally to all atoms in an optical lattice. While some of them are sitting in pairs in one of the double wells, others may remain isolated in a single subwell during the gate. The decision which and how many atoms will be paired or isolated, depends on the specific algorithm. Consequently we need to ensure that one suitable gate is applied for both paired and isolated atoms. 
To obtain the matrices $\hat{\Psi}_1(\tau)$ and $\hat{\Psi}_2(\tau)$, we apply the simulation for each state
$\psi_n \in \{ w_L, w_R\}$ or $\psi_n \in\{w_{LL}, w_{LR}, w_{RL}, w_{RR} \}$ and then calculate $\hat{\Psi}_{n,m}(\tau)=\int_{{-}\infty}^\infty \psi_m^*(x) \psi_n(x,\tau) dx$.
We then determine the gradient $\partial_{c} \epsilon={-}(\vert O_1 \vert \partial_{c} \vert O_1 \vert{+} \vert O_2 \vert \partial_{c} \vert O_2 \vert)$ of the infidelity with respect to a given control $c$ (such as the scattering length). The gradient of the state $\psi'_{c,n}:=\partial_{c}\psi_n$
can then be simulated by taking the derivative of the Schrödinger equation in \cite{Kuprov2009,Machnes2018} and by solving
\begin{equation*}
i \hbar \partial_t \psi'_{c,n} (x,t) =H(x,t) \psi'_{c,n}(x,t)
+ H'_{c} (x,t) \psi_n(x,t).
\end{equation*}
A similar equation can then be defined for the Hessian or higher derivatives of the state if needed.
If $\psi'_{c,n}$ is zero at $t=0$, we only need to simulate the state gradient during the time where $H'_{c}$ is non{-}zero. Afterwards the equation is just a normal Schrödinger
equation and we can use back propagation to calculate the gradient of the infidelity.
The only control for which the gradient is not zero at $t=0$ is the scattering length since the Wannier states at the same subwell (see Fig.~\ref{Basis Wannier states})
depend on the interaction energy. In general, to calculate the gradient of a Wannier state $w'_c$, we determine the derivative of the defining equation $\partial_c \left[(\mathrm{x}{-}x_0) w(x{-}x_0) \right]=0$ (see the beginning of Sec.~\ref{The system}) and also the derivative of the stationary Schrödinger equation $\partial_c \left[ (H(x){-}E) e^{iqx}u_q(x)\right]=0$ as $\mathrm{x}$ is given in the Bloch basis.

\subsection{Transfer function\label{sec:transfer:function}}
\begin{figure}[t]
\includegraphics[]{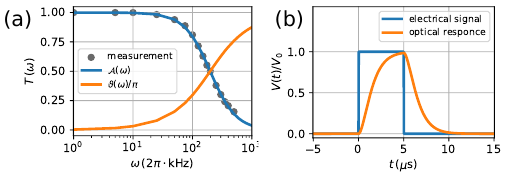}
\caption{\justifying \textbf{Transfer function for the electrical signal and its resulting optical response.} The Butterworth transfer function in (a) \cite{exp_data} yields the piece-wise constant
electrical signal block in (b) which has a delayed optical response. \label{fig:7}}
\end{figure}

The optical potential will be created in an experiment
via a piece-wise constant signal with a step size of $\Delta t{=}\SI{5}{\micro \second}$ \cite{exp_data} from an electrical signal
$V_{\mathrm{el}}(t)$ which is sent to an optical device.
In an experiment with a finite bandwidth, the optical response would no longer
be piece-wise constant.
The optical signal is calculated using the Fourier transform $\tilde{V}_{\mathrm{el}}(\omega)$
of the electrical signal $V_{\mathrm{el}}(t)$ (see, e.g., page 85 of \cite{Tang2007}), i.e.,
\begin{align*}
V_{\mathrm{el}}(t) &=\sum_{n=1}^{\tau/\Delta t} V_n \rect \left[ t{-} \left(n-\frac{1}{2} \right)\Delta t \right], \\
\tilde{V}_{\mathrm{el}}(\omega) &=\frac{\Delta t}{\sqrt{2 \pi}} \sinc \left( \frac{\omega \Delta t}{2} \right) \sum_{n=1}^{\tau/\Delta t} V_n e^{i(n{-}\frac{1}{2})\omega \Delta t}.
\end{align*}
Then one multiplies the Fourier transform of the electrical signal with the complex Butterworth transfer function
$T(\omega)=\mathcal{A}(\omega)e^{i \vartheta(\omega)}$ from the optical part of the system.
The transfer function consists of an amplitude damping $\mathcal{A}(\omega)$ as well as
a phase shift $\vartheta(\omega)$ for higher frequencies 
as detailed in Fig.~\ref{fig:7}(a).   
We then transform it back to get the optical response 
\begin{equation*}
V_{\mathrm{opt}}(t)=\frac{1}{\sqrt{2 \pi}}\int \limits_{{-}\infty}^\infty T(\omega) \tilde{V}_{\mathrm{el}}(\omega) e^{{-}i\omega t} d\omega.
\end{equation*}

To characterize the frequency response of the optical system, the experimentalists \cite{exp_data} applied a sequence of programmed piece-wise constant drive signals to an acousto-optic modulator (AOM) and recorded the resulting optical intensity using a high-bandwidth photo-detector. By varying the step duration $\Delta t$ (corresponding to drive frequencies $f=1/\Delta t$), the relative amplitude and phase response of the system were extracted and used to determine the complex transfer function $T(\omega)$. The resulting amplitude attenuation and phase delay are shown in Fig.~\ref{fig:7}(a),
and this measured transfer function is employed in the reconstruction of the optical waveform $V_{\mathrm{opt}}(t)$.
This straight-forward approach nicely works in our case to determine the transfer function. If nonlinear effects in the transfer function become relevant, further approaches are available to determine the transfer function and to apply it in the optimization \cite{Singh2023,Motzoi2011,Yang2025}.

The number of piece-wise constant control values $V_n$ in our optimization is given by $\tau/\Delta t$, while the system itself will only see the optical response obtained by applying the transfer function.
For the gradient-based optimization, one would only need to simulate the
gradient with respect to $V_n$ from $t=(n{-}1)\Delta t$ until $t=n\Delta t$
if one has an actual piece-wise constant optical response from the optical device. In our case, we need to simulate the gradient until the optical
response is zero again, which
is approximately at $t=(n{+}2)\Delta t$ as one can see in Fig.~\ref{fig:7}(b).

\section{Optimizing \texorpdfstring{$V_s(t)$}{Vs(t)}, \texorpdfstring{$V_l(t)$}{Vl(t)} and \texorpdfstring{$a_{1\mathrm{D}}$}{a1D} using a 1D-confinement simulation\label{Optimization of Vs, Vl and a in the 1D confinement}}
Before we optimize our three controls given by $V_s(t)$, $V_l(t)$, and the effective scattering length $a_{1\mathrm{D}}$ for a gate optimization, we first focus in Sec.~\ref{State{-}to{-}state preoptimization} on the single-atom case and optimize only the
intensity of the short and the long lattice for different transition ratios $\alpha$ between left and right.
We then take the optimized pulse and apply it to a simulation
of two atoms while only optimizing the scattering length for each of those pulses. 
In the second step (see Sec.~\ref{Unitary gate optimization}), we optimize all three controls again, but this time simultaneously
in a gate optimization where the previous results are taken as an initial guess.
This is motivated by the fact that the simulation time for two atoms grows quadratically
due to a quadratic growth in the
composite Hilbert space, while the simulation time for a single atom only grows
linearly.
Additionally, the optimization algorithm needs more iterations to
find a solution for the time-dependent pulses $V_s(t)$ and $V_l(t)$ as well as for the time-constant parameter $a_{1\mathrm{D}}$. Therefore, one wants to simulate only a few iterations with two atoms, especially for $V_s(t)$ and $V_l(t)$.

\begin{figure}[t]
\includegraphics[]{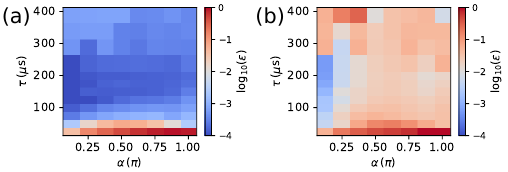}
\caption{\justifying \textbf{Infidelity of the state-to-state preoptimization.} 
Following our flow chart in Fig.~\ref{Process} (see Sec.~\ref{sec:quantum:gate:optimization}), we show the results of the
first two steps corresponding to the state-to-state optimizations 
in both subfigures:
(a)~The logarithm of the infidelity for a single-atom optimization of the lattice intensities $V_s(t)$ and $V_l(t)$. 
(b)~Same for the two-atom optimization of the effective scattering length $a_{1\mathrm{D}}$, where the time-dependent parameters are left unchanged. The infidelity for the single-atom swap is much smaller
than for an entangling collision of two atoms.\label{fig:8}}
\end{figure}

\subsection{State{-}to{-}state preoptimization\label{State{-}to{-}state preoptimization}}
We will now apply the gradient-based optimization algorithm BFGS \cite{Optim2018}, used previously in the context of the quantum optimal control method GRAPE \cite{Khaneja2005} for quasi-Newton \cite{Fouquieres2011,Motzoi2011} and Newton \cite{Dalgaard2020} optimizations. Thus, we find the best pulse for different gate times $\tau \in [\SI{25}{\micro \second}, \SI{400}{\micro \second}]$
and different transition ratios $\alpha \in [{\pi}/{8}, \pi]$ when starting with an atom in the left (L) Wannier state and moving to the right (R) Wannier state. Since we are not breaking the
symmetry of the potential during the state-to-state optimization, we do not need to include the reversed transition (from R to L). The Wannier state corresponds to an optical lattice with respective recoil energies of $V_s(0)=40 E_{r,s}$
and $V_l(0)=30 E_{r,l}$. Using the target gate $\hat{P}_{T,1}(\alpha)$ from Eq.~\eqref{eq:P:T:1},
we can define a target state as $\vert \psi_{T,1} (\alpha) \rangle=\hat{P}_{T,1}(\alpha)\vert w_L \rangle$.
In general, we define the infidelity for a state-to-state optimization 
by
\begin{equation*}
\epsilon=1-\vert \langle \psi_{T}(\alpha) \vert \psi(\tau) \rangle \vert^2,
\end{equation*}
where $\vert \psi_{T}(\alpha)\rangle $ describes a general target state.
We use this definition in this subsection and in Fig.~\ref{Case Opt} in Sec.~\ref{Case optimization}.
In Fig.~\ref{fig:8}(a),
we can see that we get quite good results for a single-atom transition
with gate times above $\tau > \SI{50}{\micro \second}$. For shorter gate times, we are only limited by technical restrictions such as the piece-wise constant steps size $\Delta t$
or the maximal recoil energies of the laser system. For two atoms as detailed in 
Fig.~\ref{fig:8}(b),
we obtain higher infidelities for the target state
$\vert \psi_{T,2} (\alpha) \rangle=\hat{P}_{T,2}(\alpha)\vert w_{L,R} \rangle$ [see Eq.~\eqref{eq:P:T:2}]
as the optimization of $V_s(t)$ and $V_l(t)$
does not consider the interaction,
as explained in Sec.~\ref{sec:quantum:gate:optimization}.
Nevertheless, this result is a good starting point for the gate optimization with one and two atoms.

\subsection{Unitary gate optimization\label{Unitary gate optimization}}
\begin{figure}[t]
\includegraphics[]{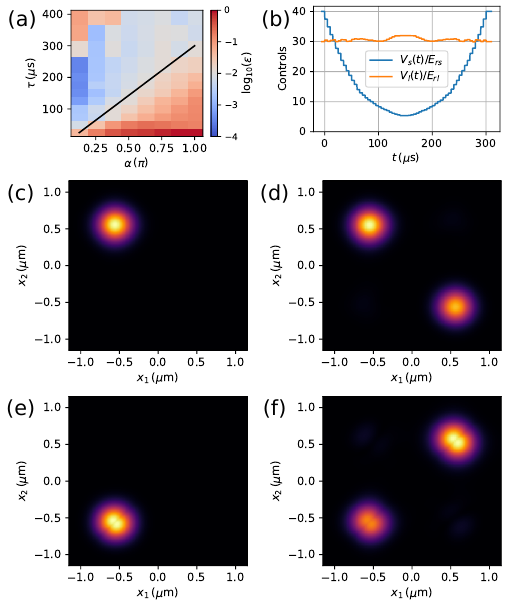}
\caption{\justifying \textbf{Optimization results.} 
Following our flow chart in Fig.~\ref{Process} (see Sec.~\ref{sec:quantum:gate:optimization}), we show the result of
the third step corresponding to the gate optimization 
in subfigure (a).
(a) Logarithm of the infidelity from the gate optimization of the lattice intensities $V_s(t)$, $V_l(t)$ and the effective
scattering length $a_{1\mathrm{D}}$. (b) Optimized pulse for $\alpha=\pi$, $\tau=\SI{300}{\micro \second}$
with an optimal scattering length of $a_{1\mathrm{D}}=-11925 a_0$. (c) and
(e) Absolute value of the two basis states $w_{LR}$ and $w_{LL}$ [where (c) agrees with Fig.~\ref{Basis Wannier states}(a)]. (d) and (f) The final states after applying the pulse
from (b) to the initial states in (c) and (e). The infidelities $\varepsilon$ are given by $0.79\%$ 
and $3.75\%$ for 
(d) and (f), respectively.}
\label{Gate Opt}
\end{figure}

The preoptimized controls for $V_s(t)$, $V_l(t)$, and the effective scattering length $a_{1\mathrm{D}}$
are now used as initial guesses for our gate optimization and this yields 
the pulse shape and the associated dynamics shown in Fig.~\ref{Gate Opt}.
The logarithm of the error plotted in Fig.~\ref{Gate Opt}(a) 
reveals a linear dependency of the gate time $\tau$ on the transition ratio $\alpha$.
The empirical quantum speed limit of the system is given by the minimal ratio
${\tau}/{\alpha}$. It is marked in Fig.~\ref{Gate Opt}(a)
with a black diagonal line and is approximately given by ${300}/{\pi}\, \SI{}{\micro \second}$.
At the empirical speed limit, we observe an error of approximately $1\%$.
Figure~\ref{Gate Opt}(b) 
shows an example of a pulse with
$\alpha=\pi$, $\tau=\SI{300}{\micro \second}$, while 
Figs.~\ref{Gate Opt}(d) and \ref{Gate Opt}(f) 
illustrate the respective simulation results 
at final time, where the goal for $\alpha=\pi$ was to end
in an equal superposition between $w_{LR}$ and $w_{RL}$ in 
Fig.~\ref{Gate Opt}(d) 
or between $w_{LL}$ and $w_{RR}$ in 
Fig.~\ref{Gate Opt}(f).
While the optimization managed to reduce the
mixture between the two cases in 
\ref{Gate Opt}(d) and \ref{Gate Opt}(f),
we have not obtained a perfectly equal superposition.
On the other hand, the effective scattering length of $a_{1\mathrm{D}}=-11925 a_0$ with a Bohr radius of
$a_0 \approx \SI{52.9}{\pico \meter}$ is too high if the atoms start in different
subwells in Fig.~\ref{Gate Opt}(c) and too low if they start in the same subwell in Fig.~\ref{Gate Opt}(e).
This notable difference in behavior for the cases of starting in different subwells or the same subwell
is the result of the momentum dependency of the interaction and is analyzed in Sec.~\ref{Case optimization}.
In addition, we now want to also estimate the corresponding real scattering length $a$ 
for the case of $\beta_{z}=10.16^\circ$
and values of $V_y=40 E_{r,y}$ and $V_z=40 E_{r,z}$, which leads to an average oscillation length $l_\perp=\SI{0.22}{\micro \meter}$ and, consequently, to an approximate value of $a=1069 a_0$. This relatively high value is caused by our experimentally oriented assumption of a small wave number $k_z$ in $z$-direction
(see Sec.~\ref{The system}). Such a high value may also affect the validity of the effective 1D s-wave scattering in Eq.~\eqref{U1D}. In contrast, if we set $\beta_{z}$ to $26.7^\circ$, we would get an average oscillation length of $l_\perp=\SI{0.10}{\micro \meter}$ and
an estimated scattering length of $a=283 a_0$, which corresponds to a typical range in experiments.

As discussed before, the most limiting factor is the piece-wise constant step size
$\Delta t=\SI{5}{\micro \second}$. It would be advantageous to reduce it at least by a factor of $2$ and increase the maximal recoil energy $E_{r,l}$ of the attractive long lattice by a factor of approximately $1.5$ if one applies gates with ${\tau}/{\alpha} \ll {300}/{\pi}\, \SI{}{\micro \second}$. Then one would have a lot more freedom for short pulses and one could probably lower the minimal ratio of ${\tau}/{\alpha}$.

\subsection{Case optimization for different initial states \label{Case optimization}}
In Sec.~\ref{Unitary gate optimization}, we optimized pulses to obtain
the unitary gates $\hat{P}_{T,1}$ and $\hat{P}_{T,2}$ for 
the one- and the two-particle basis, respectively.
This has resulted in an error of approximately $1\%$ using experimentally realistic values.
Nevertheless, we now aim at separating the optimization of the two two{-}atom cases, where we start from the same subwell ($w_{LL}$ or $w_{RR}$) or different subwells ($w_{LR}$ or $w_{RL}$) as detailed in Sec.~\ref{State{-}to{-}state preoptimization}.
This separation is motivated by the fact that some applications considers only one of these two cases, while the other one is neglected. While the applications for starting in different subwells is discussed in the context of quantum simulation and quantum computing \cite{Chalopin2025,Singh2025,Bojovic2025}, the case of starting in the same subwell has applications in quantum chemistry \cite{Gkritsis2025}.
Furthermore, the stand{-}alone optimization for the separate cases should yield much better results than the combined 
optimization in Sec.~\ref{Unitary gate optimization}.
This approach is mainly focused on the transition ratio
$\alpha=\pi$ as its final state is an equal superposition of the two possible states of the same case it started with. Numerically, the separation here means that we perform two optimizations, while calculating only the columns of $\hat{\Psi}_{n,m}(\tau)$ from Sec.~\ref{Gradient based optimization} by starting in one of the two cases and setting the columns for starting from the respective other case to zero.

\begin{figure}[t]
\includegraphics[width=0.75\linewidth]{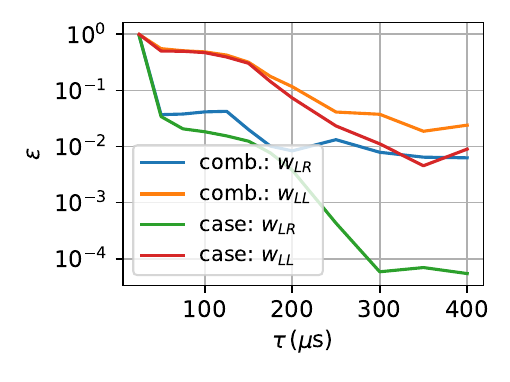}
\caption{\justifying \textbf{Infidelity of the combined optimization vs. the case optimization.} One can see a clear advantage 
for what we denote as a case optimization by performing a separate optimization for different initial states
as compared to a combined optimization for all initial states at the same time. Especially for two atoms starting at different sides of the double well ($w_{LR}$), we gain two orders of magnitude in precision for longer gate times.
Small fluctuations in the infidelity are expected as data points are obtained by
independent optimizations, which slightly differ due to, e.g., local minima.}
\label{Case Opt}
\end{figure}

Even when we optimize the two two{-}atom cases separately, we still include the one{-}atom case in both optimizations,
as our pulses need to be also applicable to single atoms in a double well.
For $\alpha=\pi$, Figure~\ref{Case Opt} compares the results 
of the combined optimization from Sec.~\ref{Unitary gate optimization} with the case
optimization from this section. In both cases, the case optimization results in a significant decrease in the infidelity. If the atoms start on different sides of the double
well, the difference is around two orders of magnitude for gate times $\tau \ge \SI{300}{\micro \second}$. 
One also sees an
advantage by separating the optimization if
the atoms start in the same subwell, even thought the difference in this case is smaller.

Why do the results of the two cases appear to be quite different? 
The answer lies in the previously mentioned momentum dependence of the interaction potential.
Two atoms starting on the same side ($w_{LL}$ or $w_{RR}$) will also move into the same direction. In other words, they will have similar momenta $q_j$.
The problem is that the interaction potential only affects the parts of the wave function which have opposite momenta, which is particularly relevant for the case of the atoms starting in opposite subwells ($w_{LR}$ or $w_{RL}$). One can
check this by taking the Fourier transformation of the pseudo{-}potential $\mathcal{F}[\delta (x_1{-}x_2)]=\delta (q_1{+}q_2)$.

Crucially, the effective interaction gets stronger
the more adiabatically the atoms are driven
from, e.g., $w_{LL}$ to $w_{RR}$. If one atom moves from $w_{LL}$ to $w_{LR}$, then the second atom needs to move from $w_{LR}$ to $w_{RR}$. 
In both of these subprocesses, the two atoms have parts of opposite momenta and consequently experience the interaction,
even if it is not as strong as in the case of moving from $w_{LR}$ to $w_{RL}$. Thus the adiabatic regime shows a reduced momentum dependency.

\section{Robustness against system impurities\label{Robustness against system impurities}}
For the optimizations in the previous sections, we have assumed perfect conditions which will not be given in realistic experiments.
We now analyze possible deviations from an idealized scenario and consider their effect on the behavior of the atoms.

\subsection{Asymmetric lattice, interaction uncertainty, and variations in the laser intensity \label{Asymmetric lattice due relative phase}}
\begin{figure}[t]
\includegraphics[]{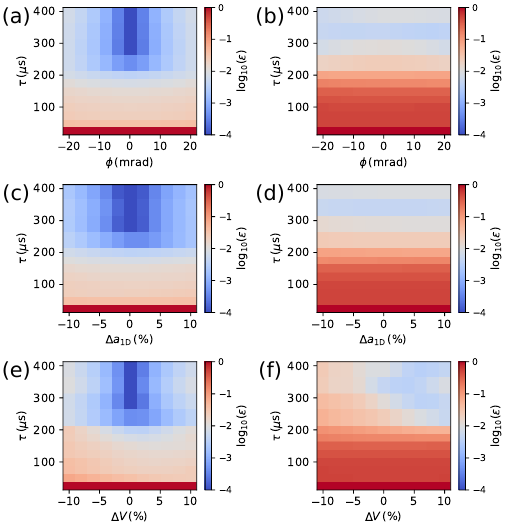}
\caption{\justifying \textbf{Robustness check.} The pulses that reach quite low infidelities in the error-free scenario
are more sensitive to
slight changes of the relative phase $\phi$, the effective scattering length $a_{1\mathrm{D}}$ and the laser intensity $V$.
This higher sensitivity can be observed for the case of two
atoms starting on different sides of the double well ($w_{LR}$) in (a), (c), and (e) when compared to two atoms starting
on the same side ($w_{LL}$) in (b), (d), and (f) for long gate times $\tau$. For shorter gate times the behavior is comparable in both cases.
The error-free scenario of $\phi=0$, $\Delta a_{1\mathrm{D}}=0$, and $\Delta V=0$ corresponds to 
the optimized results shown in Fig.~\ref{Case Opt}.}
\label{Errors}
\end{figure}

During the whole optimization process, we have assumed a perfectly symmetric lattice which corresponds to a relative phase of $\phi{=}0$ in Eq.~(\ref{eq potential}). Furthermore, we have considered a direct and simple mapping from the applied magnetic field to the effective one-dimensional scattering length $a_{1\mathrm{D}}$. Moreover, we assumed to have full control over the laser intensity. All three assumptions need to be considered carefully and we analyze the effects of the uncertainty on our optimized pulses.
We will now test our optimized pulses from Sec.~\ref{Case optimization} for different values of a time-constant non-zero relative phase and for deviations from the optimal scattering length. In Fig.~\ref{Errors}, we can see the infidelity of the case-optimized pulses from Fig.~\ref{Case Opt} in Sec.~\ref{Case optimization} for the two cases of both atoms starting on different sides, in
Figures~\ref{Errors}(a), \ref{Errors}(c), and \ref{Errors}(e), or on the same side of the double well, in  Figures~\ref{Errors}(b), \ref{Errors}(d), and \ref{Errors}(f).

In Figure~\ref{Errors},
one observes that pulses that achieve a quite low infidelity
in the error-free scenario
 are more sensitive to slight disturbances in the controls 
compared to the case with a higher infidelity.
This is observable for the case of two atoms starting on opposite sides of the double well (left column)
and gate times $\tau \ge \SI{250}{\micro \second}$, where we obtain a minimal infidelity of approximately $10^{-3}$ to $10^{-4}$. For 
gate times $\tau \leq \SI{200}{\micro \second}$, the results are more robust with an infidelity of approximately $10^{-2}$. Furthermore, one has to remark that we have used in Fig.~\ref{Errors}(b) a slightly different target state $(w_{LL}+\exp [i \gamma (\tau)] {\cdot} w_{RR})/\sqrt{2}$ with a time-dependent phase $\gamma$ as we now explain. 
Due to the relative phase of the lattice, the left and the right side of the double well do not have the same potential energy. 
Consequently, the atoms gain a time-dependent phase $\gamma$ during the operations of the gate that is roughly proportional to twice the energy difference between the states on the left and the right side of the double well.

In general, all error sources appear to have a symmetric effect on the infidelity, except for the case of two atoms starting in the same subwell in Fig.~\ref{Errors}(f). Higher laser intensity 
can lead to even better results. This is expected as we have a fixed the upper limit for the initial and final laser intensity. Thus a stronger potential will lead to a higher density in the wave functions and consequently to a higher effective interaction energy.

Nevertheless, both cases of atoms starting from the same subwell or different subwells appear to be quite robust against deviations from an idealized scenario, which is promising for an experimental realization.

\begin{figure}[t]
\includegraphics[]{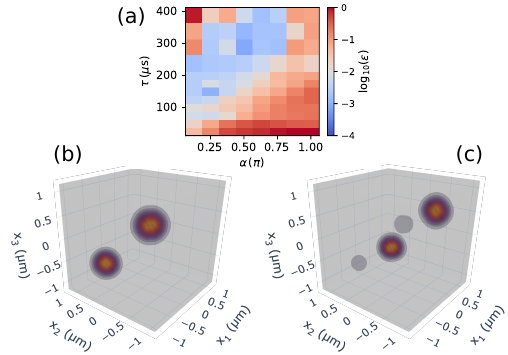}
\caption{\justifying \textbf{Three{-}atom collision.} (a) The logarithm of the infidelity of the three-atom collision in the 1D confinement. (b) The starting
state $(w_{LLR} {-} w_{LRL})/{\sqrt{2}}$ and (c) the final state for $\tau = \SI{250}{\micro \second}$ and $\alpha = \pi$ after simulating the system with the pulse from Fig.~\ref{Gate Opt}. The final state
also observes higher excitations beyond the ground states as well as a population remaining in the initial state.}
\label{three_atoms}
\end{figure}

\definecolor{myblue}{RGB}{30, 20, 160}
\definecolor{myred}{RGB}{190, 20, 20}
\newcommand{\pos}[1]{{\textcolor{myblue}{#1}}}
\newcommand{\spin}[1]{{\textcolor{myred}{#1}}}

\begin{table*}[t]
\caption{\justifying \textbf{Mapping from the total quantum state to the position-spin basis}. The total quantum state is decomposed into the position-spin basis with the help of the Slater determinant. We consider all 16 possible cases in a double well with zero to four atoms in the lowest band. This table considers a non-interacting basis, while the optimizations in this work have been performed with a basis that includes the interaction (see Fig.~\ref{Basis Wannier states}). 
When considering the interaction, one first uses this table and then afterwards formally rewrites, e.g., $\vert L \rangle \vert L \rangle$ to $\vert LL \rangle$.
Following this procedure, one obtains the correct result which is valid even though
$\vert LL \rangle \neq \vert L \rangle \vert L \rangle$.\label{table:A}}
\begin{tabular}{|c|c|}\hline
total quantum state & \pos{position}{-}\spin{spin} state \\\hline
$\vert 00 \rangle$ & $\vert \mathrm{vacuum} \rangle$ \\\hline
$\vert {\uparrow} 0 \rangle$ & $\pos{\vert L \rangle_1} \otimes \spin{\vert {\uparrow} \rangle_1}$ \\\hline
$\vert 0 {\uparrow} \rangle$ & $\pos{\vert R \rangle_1} \otimes \spin{\vert {\uparrow} \rangle_1}$ \\\hline
$\vert {\downarrow} 0 \rangle$ & $\pos{\vert L \rangle_1} \otimes \spin{\vert {\downarrow} \rangle_1}$ \\\hline
$\vert 0 {\downarrow} \rangle$ & $\pos{\vert R \rangle_1} \otimes \spin{\vert {\downarrow} \rangle_1}$ \\\hline
$\vert {\uparrow} {\downarrow} \rangle$ & $(\pos{\vert L \rangle_1 \vert R \rangle_2} \otimes \spin{\vert {\uparrow} \rangle_1 \vert {\downarrow} \rangle_2} {-} \pos{\vert R \rangle_1 \vert L \rangle_2} \otimes \spin{\vert {\downarrow} \rangle_1 \vert {\uparrow} \rangle_2})/\sqrt{2}$ \\\hline
$\vert {\downarrow} {\uparrow} \rangle$ & $(\pos{\vert R \rangle_1 \vert L \rangle_2} \otimes \spin{\vert {\uparrow} \rangle_1 \vert {\downarrow} \rangle_2} {-} \pos{\vert L \rangle_1 \vert R \rangle_2} \otimes \spin{\vert {\downarrow} \rangle_1 \vert {\uparrow} \rangle_2})/\sqrt{2}$ \\\hline
$\vert D 0 \rangle$ & $\pos{\vert L \rangle_1 \vert L \rangle_2} \otimes (\spin{\vert {\uparrow} \rangle_1 \vert {\downarrow} \rangle_2 {-} \vert {\downarrow} \rangle_1 \vert {\uparrow} \rangle_2})/\sqrt{2}$ \\\hline
$\vert 0 D \rangle$ & $\pos{\vert R \rangle_1 \vert R \rangle_2} \otimes (\spin{\vert {\uparrow} \rangle_1 \vert {\downarrow} \rangle_2 {-} \vert {\downarrow} \rangle_1 \vert {\uparrow} \rangle_2})/\sqrt{2}$ \\\hline
$\vert {\uparrow} {\uparrow} \rangle$ & $(\pos{\vert L \rangle_1 \vert R \rangle_2 {-} \vert R \rangle_1 \vert L \rangle_2}) \otimes \spin{\vert {\uparrow} \rangle_1 \vert {\uparrow} \rangle_2} /\sqrt{2}$ \\\hline
$\vert {\downarrow} {\downarrow} \rangle$ & $(\pos{\vert L \rangle_1 \vert R \rangle_2 {-} \vert R \rangle_1 \vert L \rangle_2}) \otimes \spin{\vert {\downarrow} \rangle_1 \vert {\downarrow} \rangle_2} /\sqrt{2}$\\\hline
$\vert D {\uparrow} \rangle$ & 
\begin{tabular}{c} 
$\pos{(\vert L \rangle_1 \vert R \rangle_2 {-} \vert R \rangle_1 \vert L \rangle_2)\vert L \rangle_3} \otimes \spin{\vert {\uparrow} \rangle_1 \vert {\uparrow} \rangle_2 \vert {\downarrow} \rangle_3}/\sqrt{6}$ \\ 
${-}\pos{(\vert L \rangle_1 \vert R \rangle_3 {-} \vert R \rangle_1 \vert L \rangle_3)\vert L \rangle_2} \otimes \spin{\vert {\uparrow} \rangle_1 \vert {\downarrow} \rangle_2 \vert {\uparrow} \rangle_3}/\sqrt{6}$  \\
${+}\pos{(\vert L \rangle_2 \vert R \rangle_3 {-} \vert R \rangle_2 \vert L \rangle_3)\vert L \rangle_1} \otimes \spin{\vert {\downarrow} \rangle_1 \vert {\uparrow} \rangle_2 \vert {\uparrow} \rangle_3}/\sqrt{6}$ \\
\end{tabular}  \\\hline
$\vert {\uparrow} D \rangle$ & 
\begin{tabular}{c} 
$\pos{(\vert L \rangle_1 \vert R \rangle_2 {-} \vert R \rangle_1 \vert L \rangle_2)\vert R \rangle_3} \otimes \spin{\vert {\uparrow} \rangle_1 \vert {\uparrow} \rangle_2 \vert {\downarrow} \rangle_3}/\sqrt{6}$  \\
${-}\pos{(\vert L \rangle_1 \vert R \rangle_3 {-} \vert R \rangle_1 \vert L \rangle_3)\vert R \rangle_2} \otimes \spin{\vert {\uparrow} \rangle_1 \vert {\downarrow} \rangle_2 \vert {\uparrow} \rangle_3}/\sqrt{6}$ \\
${+}\pos{(\vert L \rangle_2 \vert R \rangle_3 {-} \vert R \rangle_2 \vert L \rangle_3)\vert R \rangle_1} \otimes \spin{\vert {\downarrow} \rangle_1 \vert {\uparrow} \rangle_2 \vert {\uparrow} \rangle_3}/\sqrt{6}$ \\
\end{tabular}  \\\hline
$\vert D {\downarrow} \rangle$ & 
\begin{tabular}{c} 
$\pos{\vert L \rangle_1(\vert L \rangle_2 \vert R \rangle_3 {-} \vert R \rangle_2 \vert L \rangle_3)} \otimes \spin{\vert {\uparrow} \rangle_1 \vert {\downarrow} \rangle_2 \vert {\downarrow} \rangle_3}/\sqrt{6}$ \\
${-}\pos{\vert L \rangle_2(\vert L \rangle_1 \vert R \rangle_3 {-} \vert R \rangle_1 \vert L \rangle_3)} \otimes \spin{\vert {\downarrow} \rangle_1 \vert {\uparrow} \rangle_2 \vert {\downarrow} \rangle_3}/\sqrt{6}$  \\
${+}\pos{\vert L \rangle_3(\vert L \rangle_1 \vert R \rangle_2 {-} \vert R \rangle_1 \vert L \rangle_2)} \otimes \spin{\vert {\downarrow} \rangle_1 \vert {\downarrow} \rangle_2 \vert {\uparrow} \rangle_3}/\sqrt{6}$ \\
\end{tabular}  \\\hline
$\vert {\downarrow} D \rangle$ & 
\begin{tabular}{c}
$\pos{\vert R \rangle_1(\vert L \rangle_2 \vert R \rangle_3 {-} \vert R \rangle_2 \vert L \rangle_3)} \otimes \spin{\vert {\uparrow} \rangle_1 \vert {\downarrow} \rangle_2 \vert {\downarrow} \rangle_3}/\sqrt{6}$  \\
${-}\pos{\vert R \rangle_2(\vert L \rangle_1 \vert R \rangle_3 {-} \vert R \rangle_1 \vert L \rangle_3)} \otimes \spin{\vert {\downarrow} \rangle_1 \vert {\uparrow} \rangle_2 \vert {\downarrow} \rangle_3}/\sqrt{6}$ \\
${+}\pos{\vert R \rangle_3(\vert L \rangle_1 \vert R \rangle_2 {-} \vert R \rangle_1 \vert L \rangle_2)} \otimes \spin{\vert {\downarrow} \rangle_1 \vert {\downarrow} \rangle_2 \vert {\uparrow} \rangle_3}/\sqrt{6}$ \\
\end{tabular}  \\\hline
$\vert D D \rangle$ & 
\begin{tabular}{c} 
$\pos{(\vert L \rangle_1 \vert R \rangle_2 {-} \vert R \rangle_1 \vert L \rangle_2) (\vert L \rangle_3 \vert R \rangle_4 {-} \vert R \rangle_3 \vert L \rangle_4)} \otimes \spin{(\vert {\uparrow} \rangle_1 \vert {\uparrow} \rangle_2 \vert {\downarrow} \rangle_3 \vert {\downarrow} \rangle_4 {+}\vert {\downarrow} \rangle_1 \vert {\downarrow} \rangle_2 \vert {\uparrow} \rangle_3 \vert {\uparrow} \rangle_4)}/\sqrt{24}$ \\
${-}\pos{(\vert L \rangle_1 \vert R \rangle_3 {-} \vert R \rangle_1 \vert L \rangle_3) (\vert L \rangle_2 \vert R \rangle_4 {-} \vert R \rangle_2 \vert L \rangle_4)} \otimes \spin{(\vert {\uparrow} \rangle_1 \vert {\downarrow} \rangle_2 \vert {\uparrow} \rangle_3 \vert {\downarrow} \rangle_4 {+}\vert {\downarrow} \rangle_1 \vert {\uparrow} \rangle_2 \vert {\downarrow} \rangle_3 \vert {\uparrow} \rangle_4)}/\sqrt{24}$ \\
${+}\pos{(\vert L \rangle_1 \vert R \rangle_4 {-} \vert R \rangle_1 \vert L \rangle_4) (\vert L \rangle_2 \vert R \rangle_3 {-} \vert R \rangle_2 \vert L \rangle_3)} \otimes \spin{(\vert {\uparrow} \rangle_1 \vert {\downarrow} \rangle_2 \vert {\downarrow} \rangle_3 \vert {\uparrow} \rangle_4 {+}\vert {\downarrow} \rangle_1 \vert {\uparrow} \rangle_2 \vert {\uparrow} \rangle_3 \vert {\downarrow} \rangle_4)}/\sqrt{24}$ \\
\end{tabular}  \\\hline 
\end{tabular}
\end{table*}

\subsection{Three{-}body collisions in the 1D confinement\label{Three{-}body collision in the 1D confinement}}
Continuing the discussion on robustness, another possible error can occur during the state preparation which then yields three particles in a single double well, where the position state is $(w_{LLR} {-} w_{LRL})/{\sqrt{2}}$. Moreover, considering three spins with $s_2 {=} s_3 {=}\bar{s}_1$ (where $\bar{s}_j=\,\,\downarrow$ for $s_j=\,\,\uparrow$ and vice versa), we can then
represent the spin states $\vert D{\uparrow} \rangle$ and $\vert D {\downarrow} \rangle$, and by spatial symmetry also $\vert {\uparrow}D \rangle$ and
$\vert {\downarrow} D \rangle$, in the position basis. The three particle Hamiltonian is similar to
\begin{align*}
H(\{x_j,s_j\},t) = & \sum_{j=1}^3 {-}\frac{\hbar^2}{2m} \partial_{x_j}^2{+}V(x_j,t)  \\
& + U_{1\mathrm{D}} [\delta(x_1{-}x_2){+}\delta(x_1{-}x_3)].
\end{align*}
According to the Hubbard model, the target gate for three particles is then given by $\hat{P}_{T,1}(\alpha)\otimes\hat{\mathbb{1}}\otimes\hat{\mathbb{1}}$.

In Fig.~\ref{three_atoms}, the three-atom collision does not completely behave as expected, as we will explain now. Even when most of the population moves for $\alpha {=} \pi$ from $(w_{LLR} {-} w_{LRL})/{\sqrt{2}}$ to $(w_{RRL} {-} w_{RLR})/{\sqrt{2}}$, a non{-}negligible part results in higher excitations for gate times under $\SI{200}{\micro \second}$, or remains in the initial state for gate times over $\SI{300}{\micro \second}$.
This is consistent as we have not optimized for these error states.
But the three-atom collision is not part of our focus and it is more important to know the result after the collision than to optimize for it.
In principle, one could also simulate the case of four atoms $\vert D D \rangle$, but that system is more or less frozen due to the symmetry of the position state or in
other words it is an eigenstate of the Fermi{-}Hubbard model and the result would be trivial, see Table~\ref{table:A}. Only for shorter pulses with higher kinetic energies and strong interactions, one would get higher excitations \cite{Singh2025}.

\section{Conclusion\label{Conclusion}}
We used precise simulation and optimization methods
for two-qubit gates in ultra cold fermionic atoms using a 1D confinement and going beyond Fermi-Hubbard simulations.
The simulation time approximately grows only linearly in the system size and results in
a similar precision as for earlier approaches.
Our optimizations provide a short entangling collision gate that can be applied in the experimental setting
of \cite{Chalopin2025}.
We account for experimental upper bounds for the recoil energies in the laser system, a smallest possible time step for the piece-wise constant laser intensities, as well as a transfer function from the electric signal to the optical response.
This provides realistic predictions for experimental implementations.
We have analyzed the robustness of our pulses with regard to an asymmetric lattice due to a relative phase of the laser field, an error in the interaction energy which occurs due to uncertainties in the effective interaction, 
and state preparation errors resulting in three atoms in the same double well, where the last case can be roughly approximated
in the Fermi-Hubbard model.

Our results show a characteristic behavior of the interacting atoms depending on whether the initial state starts on the same 
or opposite sides of the double well. This presents an opportunity for optimizing the different initial states separately
providing tailored optimizations for applications in quantum chemistry \cite{Gkritsis2025} and quantum simulation and quantum computing \cite{Chalopin2025,Singh2025,Bojovic2025}.
Further improvements to control sequences for two-qubit gates 
can be realized in the future with additional numerical and feedback-based 
optimizations while adapting to specific experimental constraints.
This prepares the ground for efficient and robust gates in quantum computers and simulators with fermionic atoms.

\section*{Data availability}
The data that support the findings of this article are
openly available \cite{data2026}.

\begin{acknowledgments}
We thank Petar Bojovi{\'c} for measuring the 
transfer function \cite{exp_data},
answering many questions, and providing feedback on the manuscript. We also thank Robin Groth,
Titus Franz, Philipp Preiss, and Timon Hilker for insightful discussions about the fermionic system as well as for
providing detailed experimental parameters.
Jan Reuter would like to thank Erik Weerda, Niklas Tausendpfund, Daniel Alcalde and Matteo Rizzi for support during the initial stage of research and coding.
Last but not least, we thank Eloisa Cuestas for additional comments and a last check up.
We acknowledge support from the German Federal Ministry
of Research, Technology and Space through the funding program quantum technologies—from basic research
to market under the project FermiQP, \href{https://www.quantentechnologien.de/forschung/foerderung/quantenprozessoren-und-technologien-fuer-quantencomputer/fermiqp.html}{13N15891}.
We acknowledge funding
under Horizon Europe programme HORIZON-CL4-2022-QUANTUM-02-SGA via the project 
\href{https://doi.org/10.3030/101113690}{101113690} (PASQuanS2.1).
JS has also been supported by the German Ministry of Research, Technology and Space through
the funding program quantum technologies—from basic research to market under the project
MUNIQC-Atoms, \href{https://www.quantentechnologien.de/forschung/foerderung/quantencomputer-demonstrationsaufbauten/muniqc-atoms.html}{13N16073}.
\enlargethispage{\baselineskip}
\end{acknowledgments}

\appendix

\section{Changing the long lattice in Benchmark 1 \label{Apprendix A}}

\begin{figure}[t]
\includegraphics[width=0.75\linewidth]{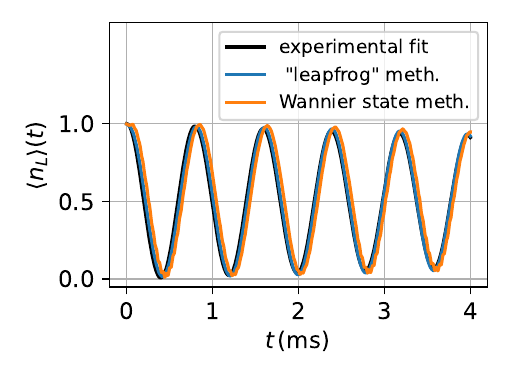}
\caption{\justifying \textbf{Variation of Benchmark 1.} 
As in Fig.~\ref{Benchmark 1},
we simulate the Rabi oscillation of a single atom with our ``leapfrog" method in blue and the Wannier-state method in orange. 
We modulate here the long lattice to obtain a relative phase as compared to modulating the short lattice
following \cite{Chalopin2025}.
As a consequence, the jitter
from Fig.~\ref{Benchmark 1} disappears.
\label{Alt Benchmark 1}}
\end{figure}

Building on the discussion of Benchmark~\ref{bench:one} in 
Sec.~\ref{Bench marking the simulation method}, we consider here an optical potential of the general form
\begin{equation}
V(x)= V_s \cos^2(k_x x {+} \phi_s) - V_l \cos^2 \left(\frac{k_x}{2} x {+} \phi_l \right)
\end{equation}
which is shifted in $x$-direction by a phase of $\phi_s$ but is almost unchanged by
the phase of $-2\phi_l$, even though both lead to the same relative phase. This can be inferred from the analytical solution of the left (L) and right (R) minima of the potential given by
\begin{equation}
x_\pm = \left[\pm \cos^{-1} \left( \frac{V_l}{4V_s} \cos \left( 2\phi_l-\phi_s \right) \right) - \phi_s \right]/k_x.
\end{equation}
In Fig.~\ref{Alt Benchmark 1}, we simulate the same scenario as in Fig.~\ref{Benchmark 1} but
the relative phase is obtained from modulating 
the long lattice (compared to the short lattice in Fig.~\ref{Benchmark 1}). We observe that the previously discussed jitter disappears.

\bibliography{Paper}

\end{document}